\documentclass[preprint2]{aastex}
\usepackage[]{natbib}
\begin{document}
\title{On QSO Distances and Lifetimes in a Local Model}
\author{M.B. Bell\altaffilmark{1}}
\altaffiltext{1}{Herzberg Institute of Astrophysics,
National
Research Council of Canada, 100 Sussex Drive, Ottawa,
ON, Canada K1A 0R6}

\begin{abstract}

It was shown previously from the redshifts and positions of the compact, high-redshift objects near the Seyfert galaxy NGC 1068 that they appear to have been ejected from the center of the galaxy in four similarly structured triplets. In this \em local \em scenario, they lie at the distance of NGC 1068, a distance much closer than a cosmological interpretation of their redshifts would imply. A large portion of their measured redshifts would then be intrinsic and it was found that this intrinsic component decreases with increasing distance from the galaxy. Here some of the consequences of assuming such a $local$ model for QSOs are examined. As has been found in several similar cases, the luminosity of the objects increases systematically with the decrease in redshift. The luminosity change cannot be Doppler related and a model in which the luminosities and intrinsic redshifts vary with time is found to fit the data best. This $local$ scenario thus appears to require a model similar to the one suggested by Narlikar and Das in which the creation of matter is ongoing throughout the life of the Universe. In fact, the observed increase in luminosity with decreasing intrinsic redshift found here, is in reasonable agreement with their prediction. In their model, matter is created with a high intrinsic redshift in mini Big Bangs and is ejected in the form of QSOs from the centers of active galaxies. From the ages of the ejection events in NGC 1068 it is found that in a relatively short time ($10^{7} - 10^{8}$ yrs) the intrinsic redshift component in these objects disappears and their luminosity approaches that of a normal galaxy. This period, which is much shorter than a Hubble time, may then determine the approximate lifetime of a QSO, and, in this model, QSOs may be the first, short-lived stage in the life of a galaxy. Perhaps of even more interest is the result that, when QSOs are assumed to be local, their generation rate is found to be constant throughout the age of the Universe. There is no need to invoke an epoch of enhanced, high-luminosity QSO production as is required in the cosmological redshift model to explain the apparent bunching-up of high-luminosity QSOs with redshifts near z = 2. Finally, because QSO lifetimes are relatively short ($<10^{8}$ yrs), an initial event (Big Bang) is still required to explain the high-redshift galaxies whose intrinsic redshift component will have long since disappeared. The Hubble expansion is therefore still expected to apply for normal galaxies.
\end{abstract}


\section{Introduction}

It has been argued \citep{arp01a,arp01b}, and references therein) that the QSO-like objects clustered near and aligned along the minor axis of active galaxies have been ejected from these same galaxies. If the ejection process is to be studied in detail, however, the positions and redshifts of a large number of objects are required. So far, redshifts are not available for most of the compact objects near active galaxies considered elsewhere. Fortunately the region around NGC 1068 has been examined very closely and most of the QSOs detected near this galaxy now have measured redshifts making it one of the best places in which to search for a possible physical association. \citet{bur99} has shown that 14 QSOs and BSOs lie within 50$\arcmin$ of this galaxy and have redshifts from 0.261 to 2.018.

\begin{figure*}
\hspace{-2.5cm}
\vspace{-3.2cm}
\epsscale{2.1}
\plotone{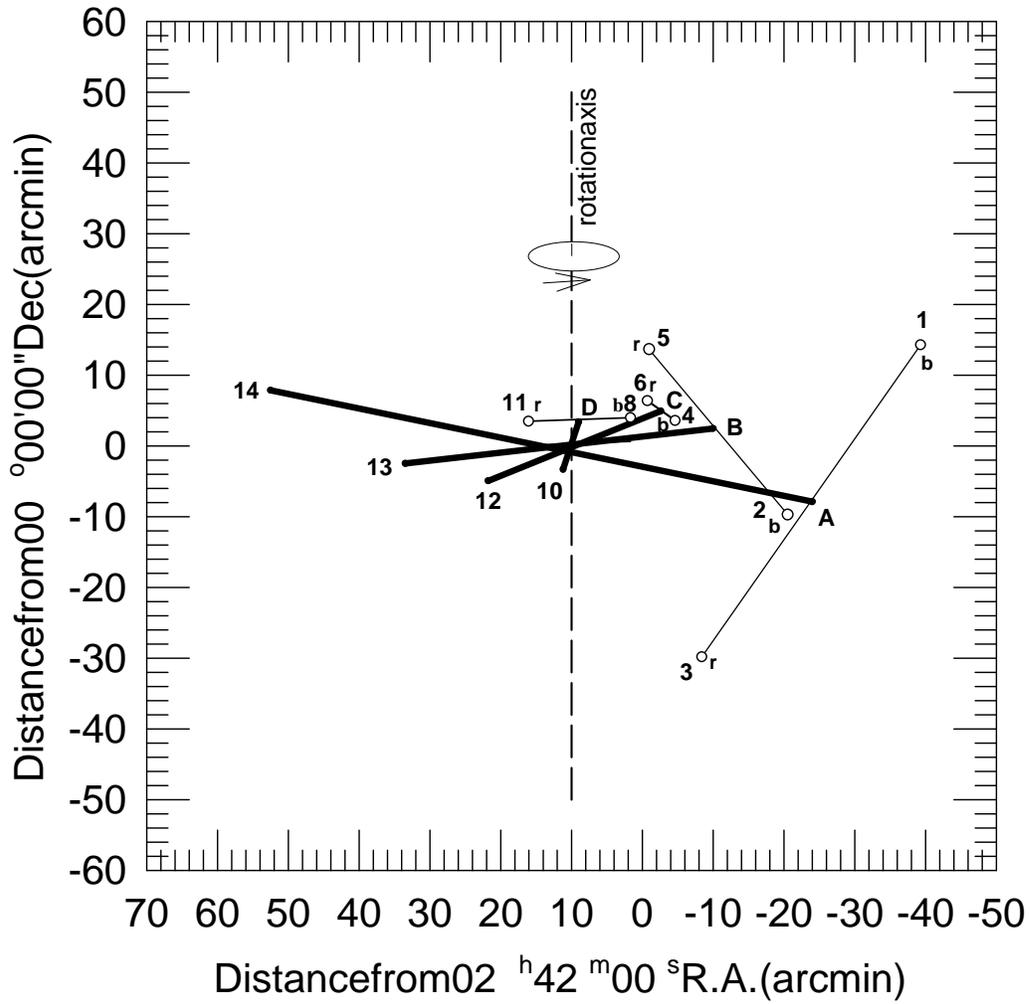}
\caption{\scriptsize{Sources plotted after removing N-S offsets equal to the displacements of their associated triplet mass centers. The letters A, B, C, and D identify the triplets. The "r" and "b" next to the paired sources indicates whether it is redshifted or blueshifted relative to the mean pair redshift. The direction of rotation of the triplet axes (heavy solid lines) is indicated at the top.\label{fig1}}}
\end{figure*}

\begin{figure}
\hspace{-2.5cm}
\vspace{-2.0cm}
\epsscale{1.1}
\plotone{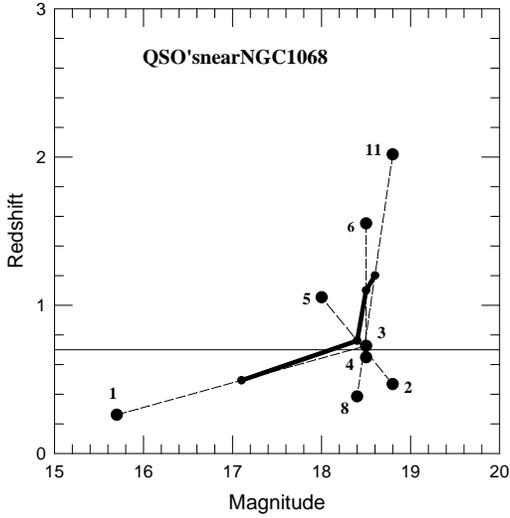}
\caption{\scriptsize{Redshift plotted versus magnitude for the paired objects. The heavy solid line shows how the mean pair redshift varies with the mean pair magnitude. Pairs are joined by dashed lines.\label{fig2}}}

\end{figure}

\begin{figure}
\hspace{-2.0cm}
\vspace{-0.5cm}
\epsscale{1.3}
\plotone{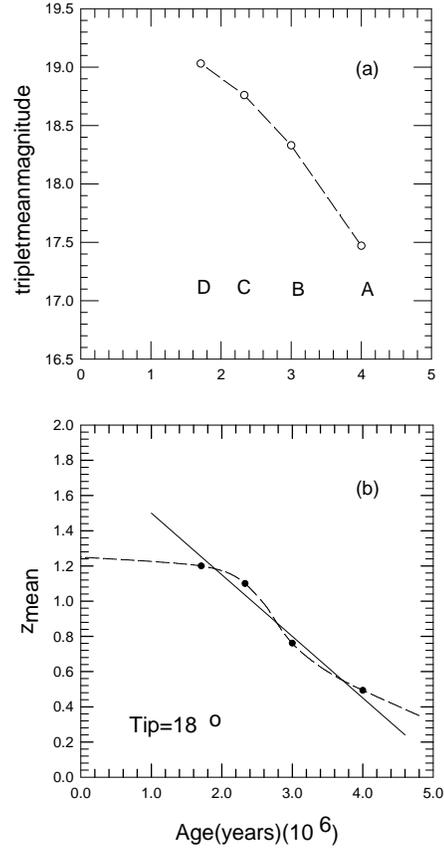}
\caption{\scriptsize{(a) Plot of mean triplet magnitude vs age. (b) Plot of z$_{\rm mean}$ vs age. The solid line assumes a linear relation. The dashed line shows qualitatively how the intrinsic redshift component might decrease if the maximum value of z$_{\rm mean}$ = 1.25 as found in paper I.\label{fig3}}}
\end{figure}

\begin{figure}
\hspace{-2.5cm}
\vspace{-2.8cm}
\epsscale{1.1}
\plotone{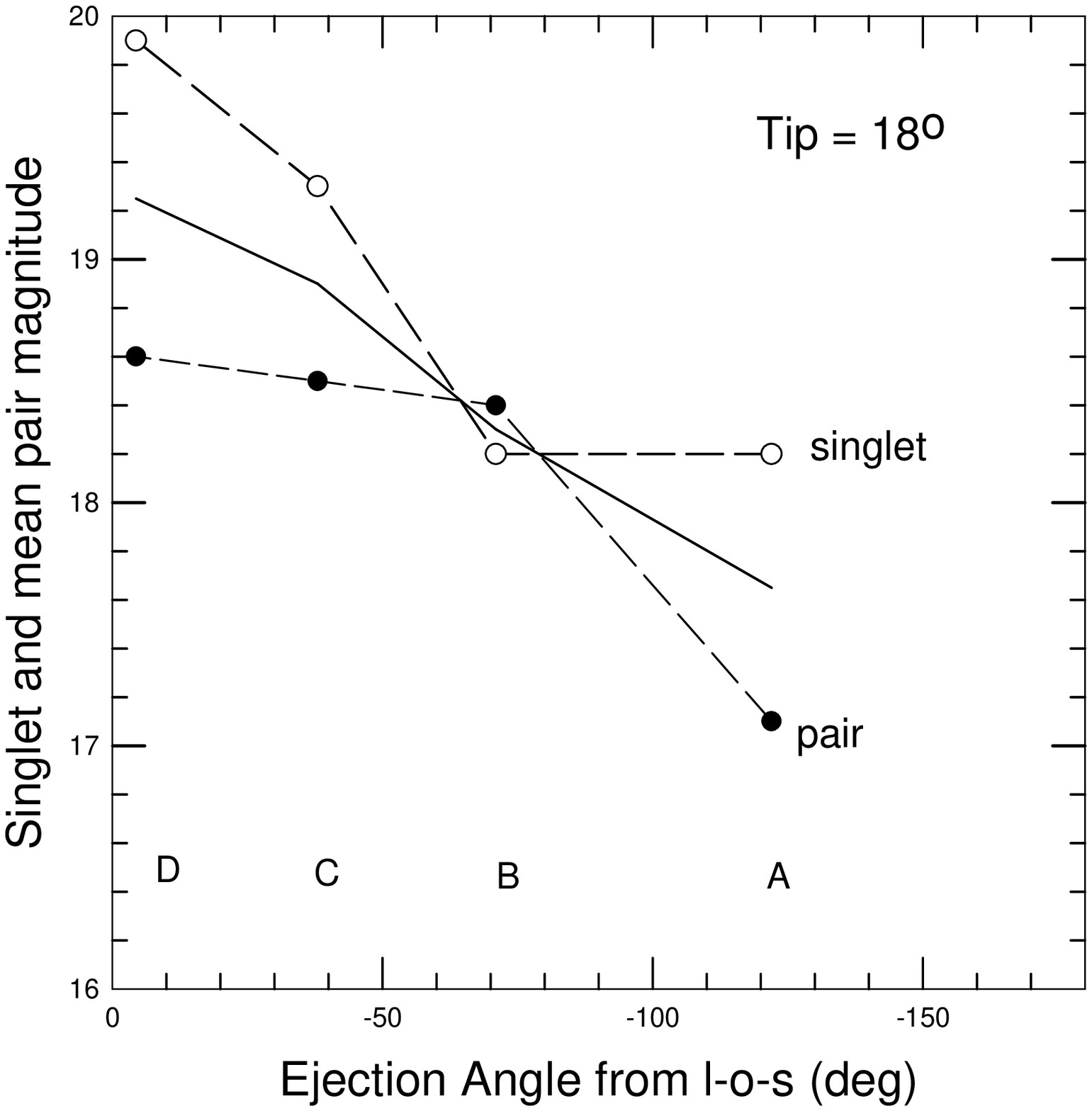}
\caption{\scriptsize{Singlet (open circles) and mean pair magnitudes (filled circles) plotted as a function of ejection angle. The solid curve is the mean of the singlet and mean pair magnitudes.\label{fig4}}}

\end{figure}

Recently, \citet{bel02} (hereafter called paper I) has shown that 12 of the compact objects clustered near NGC 1068 appear to have been ejected from the galaxy with very modest velocities (1-2$\times10^{4}$ km s$^{-1}$) in four similarly structured triplets that differ only in their size and orientation. In that study, ejection velocities and angles gave event ages $\leq6\times10^{6}$ yrs, where the event age is defined as the elapsed time since the ejection event occurred. The rotation period of the central object was also determined to be $10^{7}$ yrs, in good agreement with the rotation period of the nucleus of NGC 1068 obtained using H$_{2}$ results \citep{all01}. It was also shown that the measured redshifts are likely to be composed of several Doppler components and one non-Doppler, or intrinsic, component. 

Although source pairs have been reported previously \citep{arp97a,arp97b,arp98}, reports of triplets \citep[see][]{bur80} are less common. Here some of the consequences that result from the assumption that the QSOs near NGC 1068 have been ejected from it are examined. 

\begin{deluxetable}{cccccccc}
\tabletypesize{\scriptsize}
\tablecaption{Redshifts, magnitudes and ages of the triplets near NGC 1068. \label{tbl-1}}
\tablewidth{0pt}
\tablehead{
\colhead{Triplet(pair,singl.)}  &  \colhead{Eject. Angle($\arcdeg$)\tablenotemark{a}}  & \colhead{Mean pr. z} &  \colhead{z$_{(\rm r-b)}$} &   \colhead{Pair mag.} &   \colhead{Singl. Mag} &  \colhead{Tripl. mag} &  \colhead{Age$(10^{6})$yrs\tablenotemark{b}}
}
\startdata 
A(1-3,14)\tablenotemark{c} & -122 & 0.4935 & 0.1556 & 17.1 & 18.2 & 17.46 & 6.08 \\
B(2-5,13)  & -71  &  0.7610  & 0.1664  &  18.4  & 18.2 & 18.33 &  4.46 \\
C(4-6,12)  & -38  &  1.1005  & 0.2150  &  18.5  & 19.3 & 18.76 &  3.42 \\
D(8-11,10) & -4.4 & 1.2015   & 0.371   &  18.6  & 19.9 &  19.03 & 2.81 \\
\enddata

\tablenotetext{a}{Measured clockwise from l-o-s for $\gamma = 18^{\arcdeg}$}
\tablenotetext{b}{Assumes rotation axis tip angle $\gamma$ = $18^{\arcdeg}$ \citep{bel02}}
\tablenotetext{c}{Source numbers as listed in \citet{bur99}}
\end{deluxetable}

\section{The Data}

It was shown previously that the compact objects near NGC 1068 appear to have been ejected in triplets along the rotation axis of the central torus. The triplets are composed of a singlet and a pair that simultaneously separated in opposite directions and at 90$\arcdeg$ to the triplet ejection direction. The present locations on the sky of the pair centers and singlets are thus the result of two separate ejection components, 1) a triplet ejection component and, 2) a pair-singlet separation component. Because component 1 differs in amplitude from triplet to triplet, and can be directed in either the jet or counter-jet diresction, the current positions of the triplet centers are scattered along the rotation axis. To remove the confusion this introduces, and to make it easier to visualize the relative rotations of the pair-singlet position angles, it is of interest to see where the sources in each triplet would be located if their centers were aligned. This can be achieved simply by moving the triplet centers to a common point. In Fig. 1 the relative positions of the objects on the sky are shown after removing the N-S displacement from NGC 1068 of their associated triplet centers of mass (Fig. 4 of paper I). The triplet mass centers are now all aligned with the center of the galaxy. The triplets are labeled A to D and the numbers identify sources as listed in \citet{bur99}. The heavy solid lines connecting each singlet to each pair midpoint give the position angles of the singlet-pair separations. The change in the position angle from one triplet to the next is readily apparent and, as discussed in paper I, shows a continuous rotation from A to D, as the Seyfert galaxy rotates.

Table 1 gives the parameters found previously (paper I) for the 4 triplets. Column 1 gives the pair and singlet in each triplet, col 2 gives the ejection angle (measured clockwise from the l-o-s), col 3 gives the mean pair redshift (z$_{\rm mean}$), col 4 gives z$_{(\rm r-b)}$, defined as in Paper I, col 5, 6, and 7 give the mean magnitudes of the pairs, singlets, and triplets, and col 8 gives the triplet age. The ages of triplets C and D have been calculated here from their ejection angles using the values derived for the ages of triplets A and B in paper I, assuming a rotation axis tip angle of $\gamma$ = 18$^{\arcdeg}$ (where $\gamma$ is the rotation axis tip angle measured from the plane of the sky towards the observer at the top in Fig. 1). It has also been assumed, as in paper I, that the direction of ejection remains fixed relative to the Seyfert galaxy frame, and that the change in ejection angle is due to rotation of the central object. This is assumed to be justified by the fact that the resulting rotation period calculated for the central object is in good agreement with the rotation period determined from the results of \citet{all01}. In calculating the mean redshift only paired sources are used since the singlet redshifts have not been measured for all triplets. Although the mean pair redshifts (z$_{\rm mean}$) likely contain some Doppler components these were previously found to be small.

In fig. 2 the magnitudes of the two sources making up each pair have been plotted versus their respective measured redshifts. The numbers identify sources as listed in \citet{bur99}, and in paper I. The heavy solid line shows how z$_{\rm mean}$ varies with the mean magnitude of the paired sources. There are two things to note in this plot. The dashed lines, which connect each pair, all pass through the point near z$_{\rm mean}$ = 0.70, mag = 18.5. The mean values vary systematically from A to D, with the magnitudes becoming brighter as the z$_{\rm mean}$ redshifts decrease. The brightening varies slowly at first, but increases with age (i.e. as z$_{\rm mean}$ decreases).

\section{Magnitude and Redshift Variations with Rotation and with Time}

In Fig. 3 the mean triplet magnitudes and mean pair redshifts (cols 7 and 3 respectively, in Table 1) are plotted versus age (col 8). From Fig. 3(a), the triplets appear to be born with mean apparent magnitudes near 19.5. Their brightening gradually increases with time. In Fig. 3(b), z$_{\rm mean}$ decreases from 1.2 at age 2$\times10^{6}$ yrs to less than 0.5 at age $4\times10^{6}$ yrs. If the change is linear (solid line), the objects are born with intrinsic redshift components near z$_{\rm mean}$ = 2, and this component will have disappeared after $\sim6\times10^{6}$ yrs. If the objects are born with z$_{\rm mean}$ = 1.25 (see paper I), the change in z$_{\rm mean}$ may progress slowly at first, proceed through a more rapid stage [Age = $(2-5)\times10^{6}$ yrs], and then slow down as shown by the dashed curve. 

Since the pair-singlet position angles (col 2 of Table 1) rotate through approximately 120$\arcdeg$ from triplet A to D, both the magnitude and redshift relations in Fig. 3 approximate cosine curves. They might both then be explained by changes in the ejection angle as the galaxy rotates. However, since each singlet and pair is assumed to have been ejected in opposite directions, in a rotation model, magnitude changes with ejection angle in the singlets would be 180$\arcdeg$ out of phase with any similar magnitude changes in the pairs. In Fig. 4, where both the singlet and pair magnitudes have been plotted versus ejection angle, it is clear that \em this is not the case\em. Thus it is not possible to explain the changes in apparent magnitude using this rotation argument.

Since the ejection angle is also related to time (and the age of each triplet), the above relations can also be explained by time variations in the redshifts and magnitudes. Such a model was suggested by \citet{nar80b} who proposed that QSOs might be born out of events in galactic nucleii as matter was born out of the Big Bang. They have variable mass, increasing from zero at birth, and a high redshift that decreases with age  \citep[see][section 9]{arp98}. 

Fig. 3 has been re-plotted in Fig. 5(a) and (b), with age plotted on a logarithmic scale. Although the curves can only be estimated for ages in excess of 4$\times10^{6}$ yrs, they imply that by $\sim10^{8}$ yrs the apparent magnitudes will have approached those expected for normal galaxies at the distance of NGC 1068, and the intrinsic redshift component will have largely disappeared. This indicates that the intrinsic redshift component is a transient one, and therefore the QSOs themselves, are short-lived ($10^{7} - 10^{8}$ yrs) compared to the age of the Universe ($\sim1.4\times10^{10}$ yrs). On the other hand, the rate of change of the redshift would only be $\Delta z \sim2\times10^{-6}$ yr$^{-1}$, which would be impossible to detect. Compact objects reported close to nearby active galaxies \citep{arp99b,bur99,chu98} have measured redshifts z $\lesssim2$. This may be an indication of their maximum intrinsic redshift at birth, or it may simply indicate that the objects with higher redshifts are too faint to be detected. Whatever the reason, Fig. 5 shows that this intrinsic component will have largely disappeared after $10^{8}$ yrs.

\begin{figure}
\hspace{-1.5cm}
\vspace{-1.5cm}
\epsscale{1.1}
\plotone{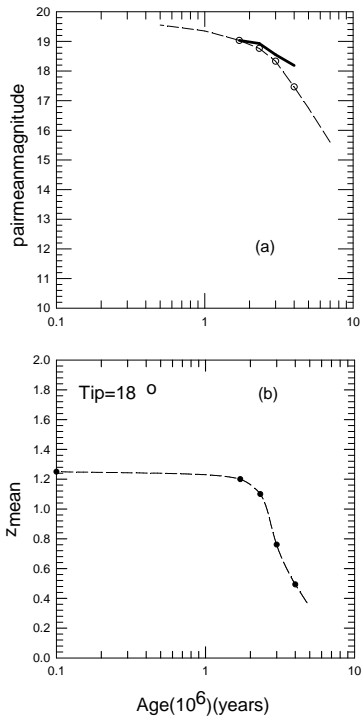}
\caption{\scriptsize{(a)Plot as in Fig. 3(a) except here age is plotted on a logarithmic scale. See text for an explanation of the heavy line. (b)Fig. 3(b) with age plotted on a logarithmic scale. \label{fig5}}}
\end{figure}

If the intrinsic redshift component disappears in $10^{7}$ to $10^{8}$ yrs, and the objects are no longer recognized as QSOs, what happens to them? The most logical assumption would seem to be that they evolve into galaxies. During their QSO phase they are several magnitudes fainter than the objects into which they would appear to evolve. If they evolve into companion galaxies, as postulated by \citet{arp98}, then in these ejections from active galaxies we are, in reality, observing a "pregnant" mother galaxy giving birth. 

\begin{figure}
\hspace{-1.0cm}
\vspace{-2.0cm}
\epsscale{1.0}
\plotone{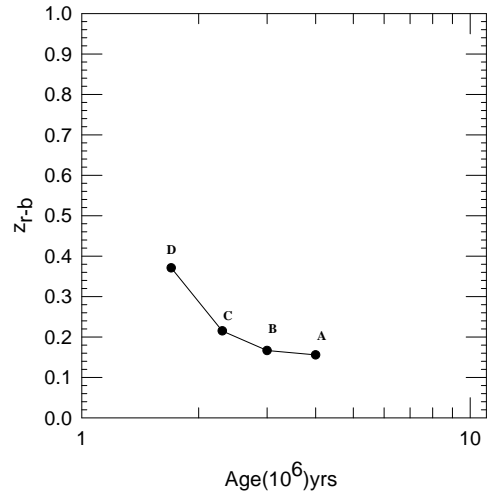}
\caption{\scriptsize{Decrease in z$_{\rm r-b}$ with time. Z$_{\rm r-b}$ is assumed to be a Doppler-related redshift. Letters identify the triplets.\label{fig6}}}
\end{figure}

The amount of time spent in the QSO stage, compared to the lifetime of a galaxy, should give information on the relative numbers of these objects. If normal galaxies are $\sim1.3\times10^{10}$ yrs old and QSOs are only around for $\sim10^{7} - \sim10^{8}$ yrs, this implies a galaxy/QSO ratio between $10^{2}$ and $10^{3}$. 

\begin{figure}
\hspace{-1.0cm}
\vspace{-2.0cm}
\epsscale{1.0}
\plotone{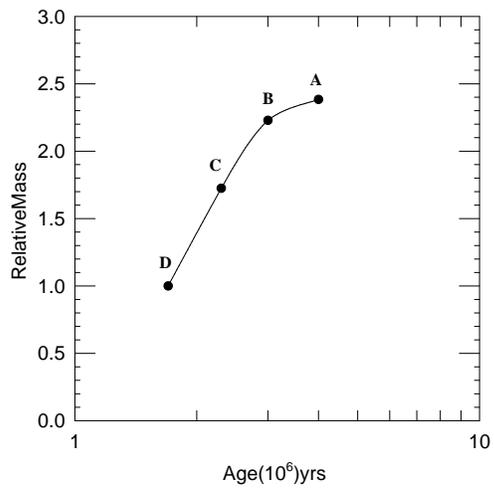}
\caption{\scriptsize{Plot of mass change with time where mass has been determined simply by assuming that momentum is conserved.\label{fig7}}}

\end{figure}

\section{High ejection velocities but low cluster velocity dispersion}

Since normal galaxies are observed in clusters, if the ejected objects evolve into normal galaxies they might be expected to remain bound to the parent galaxy. Although the ejection velocities determined for the objects near NGC 1068 are low ($\sim2\times10^{4}$ km s$^{-1}$), they still exceed the escape velocity. This may not be a serious problem, however, since, \em if momentum is conserved, \em their velocities would be expected to decrease as their masses increase with time. In fact, there is evidence for this in Fig. 6 where the residual orbital velocities (z$_{\rm r-b})$ of the paired sources (see Paper I) decrease smoothly with time. If the objects do remain in clusters, and this is far from certain, this argument might explain how the velocity might be  reduced. Unfortunately there is no way of knowing how the mass increase occurs (i.e whether it actually changes with time, as suggested by the theory of \citet[see section 5 below]{nar80b}, or whether it is simply accumulated with time, or perhaps even created directly from energy).

If the velocity change in Fig. 6 is due to increasing mass, it also implies that the radial ejection velocities will decrease with time. Since constant radial velocities were assumed in the calculation of the ages of each triplet, this would mean that the initial ejection velocities were higher than presently measured and ages would be slightly lower than calculated. Before this can be confirmed, redshifts for objects 10 and 12 will need to be measured.

\section{Tests of the Narlikar and Das Model}

\citet{nar80b} suggested several tests that might be used to check their theory. One of these predicted that the redshift dependence of particle masses in a QSO should show up in its luminosity. If the synchrotron mechanism is responsible for emission, the luminosity should relate to the QSO intrinsic redshift (z$_{\rm Q}$) by the factor  (1 + z$_{\rm Q})^{2}$ \citep{nar80b}. In Fig. 5(a) the heavy solid line shows how the apparent magnitude would be expected to change with redshift for this relation. The values have been normalized to z$_{\rm mean}$ = 1.2. Although the luminosity appears to increase more slowly than predicted by Narlikar and Das this can be explained. If it is assumed that their luminosity change occurs in the developing nebulosity or "host" galaxy, there would presumably need to be a correction required to compensate for the simultaneous dimming of the central quasar.

In the Narlikar and Das theory the manner in which mass changes with age (intrinsic redshift) is also given. As an example they point out that if a QSO of redshift z$_{\rm Q}$ = 2 is seen in association with a galaxy of redshift z$_{\rm G}$ = 0.005, then the electron mass in the latter is predicted to be 3 times the electron mass in the former. In the case of NGC 1068, if the redshift in Fig. 6 is Doppler related, as concluded in Paper I, and its decrease with time is due to increasing mass, a crude estimate of the magnitude of this mass increase might be determined by making the simple assumption that momentum is conserved. After making this assumption, the relative mass as a function of time is plotted in Fig. 7 and, as above, is in reasonable agreement with what has been predicted by the Narlikar and Das model.

\section{Logz-m$_{\rm v}$ plot.}

By late 1986, redshifts and magnitudes were known for 3594 QSOs \citep{hew87}. There are now several thousand more, many with redshifts between 4 and 5 \citep{and01}. Some of these have been plotted in Fig. 8 using the results from several different published samples. The location of strong radio galaxies on this plot \citep{san65} is indicated by the lower dashed line with a slope of 5 showing how the apparent magnitude of a "standard candle" will become fainter with increasing cosmological redshift. The vertical dotted line gives the approximate plate cut-off when many of the early surveys were carried out. Sources above logcz $\sim6$ are high-redshift objects detected in the Sloan digital Sky Survey (SDSS) commissioning data \citep{and01} and clearly have a cut-off closer to an apparent magnitude of 21. 

\begin{figure*}
\hspace{-3.0cm}
\vspace{-5.0cm}
\epsscale{2.2}
\plotone{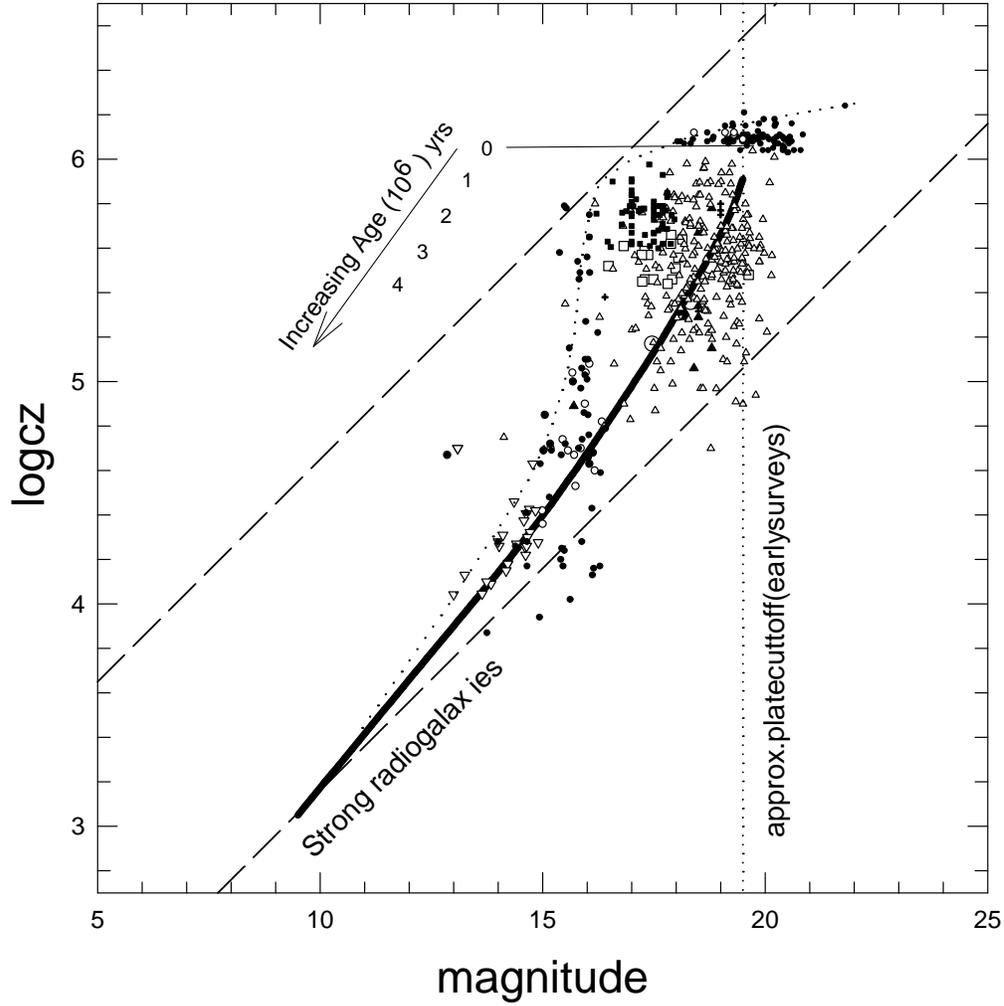}
\caption{\scriptsize{Logcz-m$_{\rm v}$ plot for QSOs from various samples. The lower dashed line indicates where strong radio galaxies are located (Sandage, 1965). Vertical dotted line indicates the approximate plate cut-off in early QSO surveys. The dotted curve indicates the approximate low-magnitude edge to the distribution. The heavy black curve indicates how QSOs evolve with time in the "local" model. Data are as follows: (inverted triangles) from \citet[Table 7]{arp90}; (upright triangles) from \citet{bec01}; (large open circles) represent mean data for sources near NGC 1068 from this paper; (filled circles with magnitudes $< 17$) from \citet{sch83}; (filled circles with magnitudes $> 17$) from \citet{and01,fan01}; (filled squares) from \citet[Table 9a]{arp90}; (filled triangles) from \citet{bur99}; (open squares) from \citet[Table 1]{fan01}; (crosses) from \citet{arp99a}; (small open circles with logcz $> 6$) from \citet[Table 2]{ken95}; (small open circles with logcz $< 5$) from \citet{lao97}.\label{fig8}}}
\end{figure*}

\begin{figure*}
\hspace{-3.0cm}
\vspace{-4.5cm}
\epsscale{2.2}
\plotone{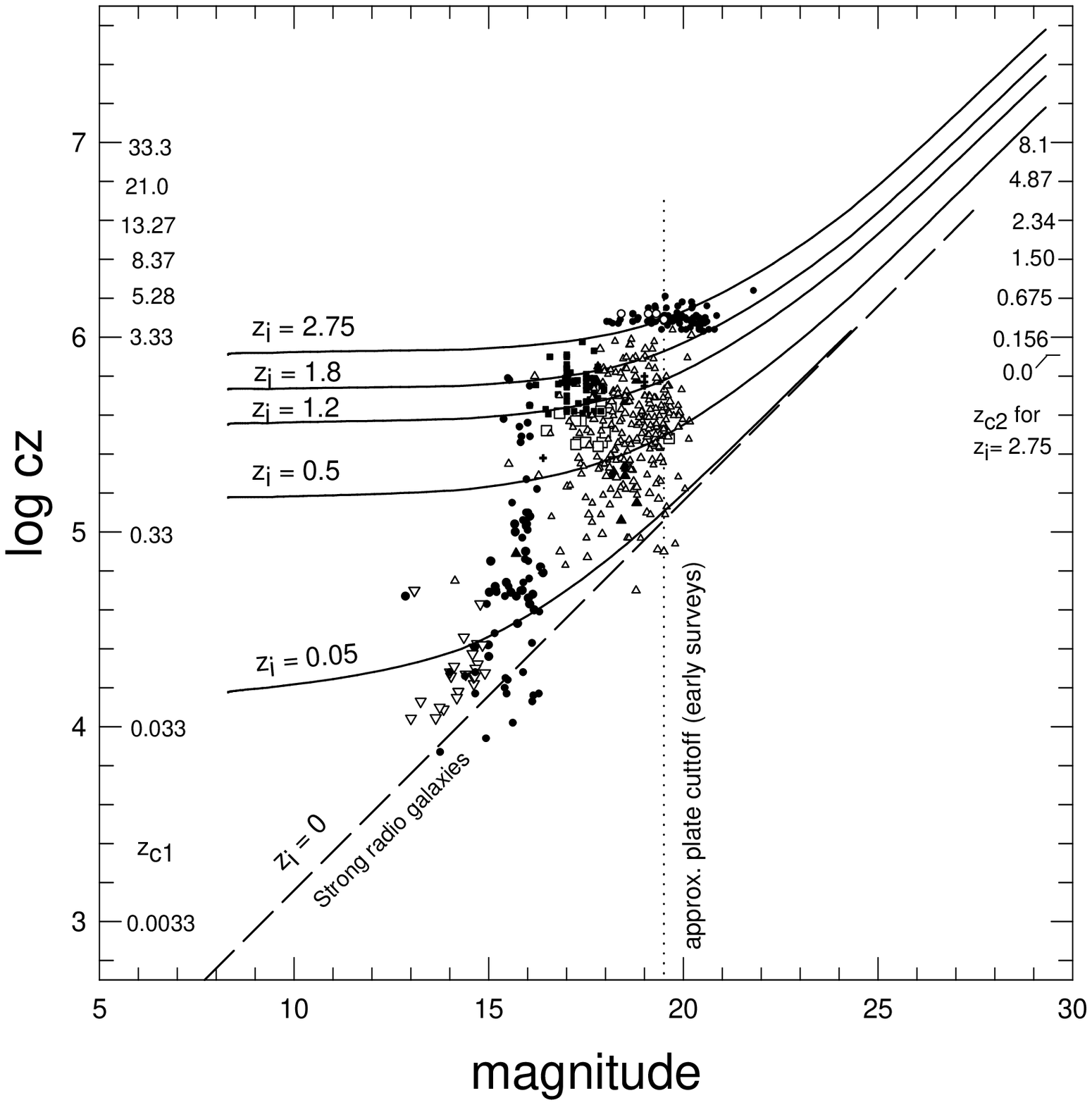}
\caption{\scriptsize{Data as in Fig. 8. Solid curves show how standard candles are expected to track if they contain different intrinsic redshift components in addition to the Hubble expansion redshift.\label{fig9}}}
\end{figure*}

For logcz values greater than 5 the low-luminosity edge has no physical meaning for this distribution other than defining the sensitivity limit. The high-luminosity edge defined roughly by the dotted curve may have real significance, however, since in most samples brighter sources would not have been missed. This is especially true in the case of the strong quasars (filled dots with mag brighter than 16.5) where the whole sky has already been sampled \citep{sch83}. It is not yet clear whether the high-luminosity edge obtained for high-redshift sources from the SDSS (logcz $>6.1$) is a real source luminosity limit or an observing limit, although it seems unlikely that it would be the latter. It is assumed here that the triangular region currently devoid of sources, having magnitudes fainter than 20 and defined by the lower dashed line and the dotted curve, will fill up with sources when adequate sensitivity is achieved. In fact, many faint galaxies with redshifts between 0.33 and 5.3 have recently been reported in the Hubble Deep Field North \citep{daw01}, and QSOs lying in this portion of the plot in Fig. 8 may already have been detected.

The data in Fig. 8 have been selected from several different samples and several of these use slightly different ways of estimating magnitudes. However, the above conclusions relating to the distribution edge defined by the dotted curve basically involve only two surveys and thus the conclusions should not be compromised by magnitude uncertainties.

In the cosmological redshift hypothesis, where the measured redshifts are assumed to be an indication of distance, the high-luminosity edge defined by the dotted curve in Fig. 8 can be explained if the highest luminosity QSOs were produced only at an epoch near z$_{\rm c1} \sim2$ (were z$_{\rm c1}$ is the measured redshift defined as a distance redshift). If similar-luminosity sources had been produced at other epochs, the dotted curve would be expected to reach the upper dashed line over the entire magnitude range from m $\sim13$ to m $\sim18$.

\begin{figure}
\hspace{-2.0cm}
\vspace{-2.0cm}
\epsscale{1.1}
\plotone{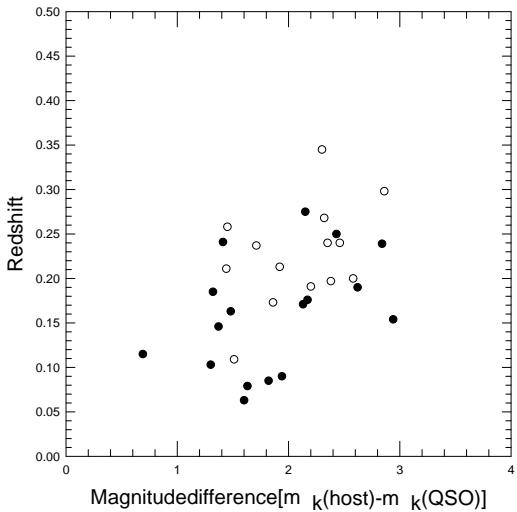}
\caption{\scriptsize{K-magnitude differences between QSOs and their host galaxies plotted vs redshift. Data are from \citet{dun93}. Filled circles are for radio-loud quasars and open circles are for radio-quiet ones.\label{fig10}}}

\end{figure}

\begin{figure}
\hspace{-2.0cm}
\vspace{-2.3cm}
\epsscale{1.1}
\plotone{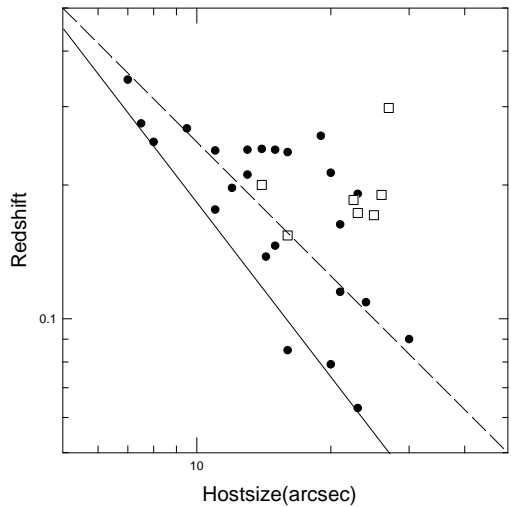}
\caption{\scriptsize{Host size after QSO subtraction, plotted vs redshift for data in Fig. 10. The open squares identify those objects with associated structure on the small chance that it might have influenced the subtraction process. The solid line shows the slope of the minimum host size. The dashed line has a slope of one.\label{fig11}}}

\end{figure}

Also included in Fig. 8 is a qualitative indication of how, in the $local$ model, the QSOs near NGC 1068 are expected to evolve (heavy solid line) from their birth with an apparent magnitude near 19.5 and intrinsic redshift near z$_{\rm i}= 2$, to magnitudes and intrinsic redshifts near 9.5 and 0 respectively. Their approximate ages enroute are indicated in the top left of the figure. In fact, there may be evidence that some normal galaxies still contain a small intrinsic redshift component \citep{rus02}. 

It is now clear from Fig. 8 that in the $local$ model, where the high-redshift QSOs near NGC 1068 are assumed to lie at the same distance as the Seyfert galaxy, that at least some of the luminous-appearing objects near the z$_{\rm c1}$ = 2 epoch are actually subluminous. It is only due to the fact that their redshifts contain an intrinsic component that they appear on the plot to be superluminous. \em Is it possible then, that the well-known bunching up of apparently luminous QSOs near z$_{\rm c1}$ = 2 \cite[see][fig. 5]{bur01} is caused entirely by the fact that the redshifts of these sources contain an intrinsic redshift component? \em

To examine this further, in Fig. 9, curves have been included showing how standard candles would track on the logcz-m$_{\rm v}$ plot in Fig. 8 if their measured redshifts contain an intrinsic redshift component (z$_{\rm i}$). Distance redshifts in the cosmological interpretation (z$_{\rm c1}$) are indicated on the left in Fig. 9 and distance redshifts in the local interpretation (z$_{\rm c2}$) are indicated on the right for sources with an intrinsic redshift component of z$_{\rm i}$ = 2.75. Distance redshifts corresponding to other intrinsic redshift values (z$_{\rm i}$) can be easily calculated if desired. Perhaps the most signifcant observation to be made from this plot is the fact that the distribution is fitted well for intrinsic redshift values z$_{\rm i} \lesssim2$, which is in excellent agreement with the range of intrinsic redshift values observed for QSOs clustered near active galaxies \citep{arp83,chu98,bur99}. In other words, \em the current logcz - m$_{\rm v}$ distribution of sources can be fitted exactly if the short-lived QSOs have intrinsic redshifts z$_{\rm i} \lesssim2$ (identical to those found for sources clustered near active galaxies), together with a distance redshift component defined by the Hubble relation. \em

The model of \citet{nar80a} does provide an alternative explanation for the Hubble expansion redshift. However, if QSO production is ongoing, and they evolve relatively quickly ($10^{8}$ yrs) into something different, the distribution in Fig. 9 appears to still require an initial Big Bang and subsequent expansion. These $local$ model results thus suggest that the QSO phase, in which intrinsic redshifts and absolute luminosities behave as predicted by the theory of Narlikar and Das, is a transient one and that for most of their lives these objects obey normal physical laws. 
It is interesting to speculate whether after the initial Big Bang matter first appeared in clumps as it does here after the "mini Bangs". If so, galaxies would be expected to appear within $10^{8}$ yrs of the Big Bang. The fact that they are initially sub-luminous by several magnitudes would prevent much earlier detection. It is important to remember that the high-redshift QSOs detected with z$_{\rm c1} > 5$ (logcz$_{c1} > 6.17$) appear to be born soon after the Big Bang only if the cosmological redshift model is assumed.
As can be seen in Fig. 9, in the $local$ model these objects have much smaller distance redshift components. 

Although it is impossible to know how much dispersion may be present in the magnitudes at a given redshift, or how much of the measured redshift is Doppler-related, broadly speaking, in Fig. 9 distance decreases \em horizontally \em to the left for sources with z$_{\rm i} > 0.1$ and magnitudes brighter than 18. \em This, then, explains the vertical appearance found for the Palomar Bright Quasar distribution (filled dots with m$_{\rm v} < 17$) from \citet{sch83}. \em Where this vertical-lying edge at magnitudes near 15.5 is interpreted as a \em high-luminosity edge \em in the cosmological interpretation, in the local model it is interpreted as a \em nearby cut-off \em that arises simply because of the rapid fall-off in source numbers as the volume of space sampled decreases.

Unfortunately, in the $local$ model, since these objects are intrinsically fainter than normal galaxies when they first appear, young ones will never be detectable to distances as great as the most distant galaxies. Thus, in this model, normal galaxies are the best means by which to sample the distant, early universe. It should also be noted that when more distant QSOs are discovered (i.e. with z$_{\rm c2}$-values $> 0.6$, logcz$_{c1} > 6.2$, m$_{\rm v} > 20$) their distribution will be expected to fall within the area tracked by the solid curves in Fig. 9 for magnitudes greater than 20.

\section{QSO Host Galaxies}

If QSOs do evolve into galaxies, finding similar-redshift "host" galaxies associated with QSOs would be expected, at least towards the end of their evolution when their intrinsic redshift component has decreased significantly. Therefore such detections in low-redshift QSOs should not be taken as proof that QSO redshifts are purely cosmological. For example, the quasars (z $< 0.6$) observed in recent searches for quasar host galaxies \citep{dun93,hoo97,mar01} would, in the $local$ model, have intrinsic redshifts z$_{\rm i}\lesssim0.5$, indicating from Fig. 5(b), that their ages would be between $4\times10^{6}$ and $10^{7}$ yrs. At this point in their evolution the presence of an evolving nebulosity would be expected. In Fig. 3 of \citet{mar01} the largest associated structure tends to be associated with the objects with lower redshifts. In the cosmological interpretation it might be argued that this is because the objects are closer. In the $local$ model it can be argued that they are simply more evolved.

\citet{dun93} examined 32 radio loud and radio quiet quasars for the presence of a host galaxy. Their analysis shows several relations that support the cosmological redshift interpretation. However, here again, equally valid arguments can be made in support of the $local$ model. For example, in Fig. 10 the difference between the K magnitude found for each QSO and that of its associated host is plotted vs redshift. Radio loud sources are plotted as filled circles and radio quiet ones as open circles. The figure shows a clear trend in both samples for this difference to increase with redshift. In the $local$ model, where z is assumed to be largely intrinsic, this can be explained by an increase in the luminosity of the host galaxy as z decreases, accompanied by a simultaneous decrease in the QSO luminosity. However, a similar effect might be expected in the cosmological model if the size of the QSO subtracted (12 arcsec aperture used) was constant for all redshifts. On the other hand, if the subtraction technique did effect the magnitude of the host, their claim that the resulting logz-m$_{\rm K}$ plot for the host galaxies being good "standard candles" would no longer be valid.

In Fig. 11, the size of the host galaxy is plotted vs redshift. Here the size was measured directly from Fig. 5 of \citet{dun93} and represents the maximum dimension of the host galaxy after subtraction of the QSO. Although there may be a limit to the minimum size of the host, that decreases with increasing redshift (solid line) there otherwise does not appear to be a significant correlation between size and redshift for the host galaxy. Although an increase in size with decreasing redshift is predicted in both models, the lack of a significant correlation in the $local$ model can be explained if the redshifts are due to a combination of distance and intrinsic redshifts and not predominantly intrinsic as assumed.

It is concluded here that for low-redshift QSOs the presence of an associated host galaxy is not a strong argument in favor of either model.

\section{Where are the objects of intermediate age?}

It has been shown that, in the $local$ model, the lifetime of a QSO is relatively short, lasting less than 10$^{8}$ yrs. \citet{san65} obtained similar results for QSOs in the cosmological model. \em In both models a QSO with a redshift z = 0.1 must be less than 10$^{8}$ years old and therefore was born less than $10^{8}$ years ago. \em QSOs are observed with measured redshifts that cover the entire range of redshifts from z $\sim0.1$ to z $> 5$. In the cosmological redshift model, these short-lived objects thus were being made within 1 Gyr of the Big Bang and have continued to be produced throughout the last 12-13 Gyr. In the $local$ model, because they are born intrinsically fainter, they are visible only at more recent epochs, however, their production is also assumed to have been continuous since the Big Bang. If it is assumed that these objects don't just disappear, they must still exist in some form, and \em in both models they must then also exist as objects whose ages are continuous from $10^{8}$ to $1.3\times10^{10}$ yrs. \em Most normal galaxies are assumed to have ages greater than 10 Gyr. Have the galaxies with ages near 1 Gyr been somehow overlooked? Are they yet to be discovered? How accurate are the ages that have been determined? These are only some of the questions that need to be asked if QSOs are a short-lived phenomenon, and they need to be asked regardless of whether the cosmological model or the $local$ model is the correct one. 

\section{Conclusions}

Using the results found previously for the QSOs near NGC 1068 \citep{bel02} it is shown that, in this $local$ model, changes in their magnitudes and redshifts with time during the QSO stage can be fitted best to a model similar to that proposed by \citet{nar80b}. The compact objects near NGC 1068 (z$_{\rm c}$ = 0.0038) are born with apparent magnitudes near 19.5 (or fainter) and a large intrinsic redshift component that decreases as their luminosity increases. After $\sim10^{8}$ yrs this intrinsic redshift component will have largely disappeared and they are assumed to evolve into galaxies. In this scenario, the formation of QSOs (and therefore also galaxies) is \em continuous \em and \em uniform \em throughout the entire age of the Universe. These results (varying luminosity, redshift and mass) differ from the model of Narlikar and Das in that they seem to apply only during the object's birth and the first brief period thereafter (the QSO stage). 
Because the QSO stage is so short, an initial Big Bang still appears to be required to explain the existence of high-redshift galaxies whose intrinsic redshift component should have long ago fallen to near-zero values. It has been shown that in this model it is no longer necessary to hypothesize a period of high-luminosity QSO production near z = 2, as is required for the cosmological redshift model. In the QSO stage, the change of luminosity with intrinsic redshift and the change of mass with time are both found to agree well with the predictions of the Narlikar and Das theory. 
The amount of time spent in the QSO stage ($\sim10^{8}$ yrs) compared to the life of a galaxy, results in a galaxy/QSO number ratio of $\sim10^{2} - \sim10^{3}$. 

Finally, it is suggested that, in the future, if this $local$ model is ever to be convincingly confirmed, or proven to be incorrect, it is very important to continue to measure the redshifts and magnitudes of those QSOs that appear to have been ejected from nearby active galaxies. This is especially true for those cases where a large number of objects have been reported.

\end{document}